\documentclass[conference]{IEEEtran}
\IEEEoverridecommandlockouts

\usepackage{cite}
\usepackage{amsmath,amssymb,amsfonts}
\usepackage{algorithmic}
\usepackage{graphicx}
\usepackage{textcomp}
\usepackage{xcolor}
\usepackage{url}
\usepackage{subfig}
\usepackage[bookmarks=false,hidelinks]{hyperref}
\usepackage{tabularx}
\usepackage{array}
\usepackage{makecell}

\newcolumntype{Y}{>{\raggedright\arraybackslash}X}

\newcommand{\BibTeX}{{\rm B\kern-.05em{\sc i\kern-.025em b}\kern-.08em
    T\kern-.1667em\lower.7ex\hbox{E}\kern-.125emX}}

\begin{document}

\title{WiWorld-RealData: A Real-World Multi-Modal Dataset for 6G Wireless World Models}
\author{
\IEEEauthorblockN{
Yinyin Jiao$^{1}$, Huixin Xu$^{1}$, Jianhua Zhang$^{1}$, Yuelong Qiu$^{1}$, Jingjing Wang$^{1}$
}
\IEEEauthorblockN{
Shaoyi Liu$^{1}$, Yuxiang Zhang$^{1}$, Li Yu$^{1}$, Xuebin Sun$^{1}$, Guangyi Liu$^{2}$
}
\IEEEauthorblockA{
$^{1}$State Key Laboratory of Networking and Switching Technology,\\
Beijing University of Posts and Telecommunications, Beijing 100876, China
}
\IEEEauthorblockA{
$^{2}$China Mobile Research Institute, China
}
\IEEEauthorblockA{
Emails: \{jyy19996289849, xuhuixin, jhzhang, yl\_qiu, wangjingjing123, sy\_liu, zhangyx, li.yu, sunxuebin\}@bupt.edu.cn,\\
liuguangyi@chinamobile.com
}
}

\maketitle

\begin{abstract}
As sixth-generation wireless systems evolve from reliable connectivity toward environment intelligence, wireless world models aim to learn how physical environments and user states affect wireless propagation, requiring real-world data with explicit correspondences between channel responses and environment observations. However, existing channel--environment datasets are predominantly simulation-based or designed for specific communication tasks, limiting their support for general environment--channel relationship learning. To address this gap, we construct WiWorld-RealData, a real-world multi-band channel and multi-modal environment sensing dataset for 6G wireless world model research. It provides synchronized channel impulse responses measured at 3.7 and 6.775~GHz together with multi-view and panoramic images, light detection and ranging point clouds, millimeter-wave radar observations, and global navigation satellite system trajectories. Unified timestamps, sample identifiers, and metadata establish sample-level correspondences across these heterogeneous modalities. The overall measurement campaign produced approximately 10~TB of data, while the current public release provides aligned channel--environment samples from a representative continuous outdoor route. A path-loss prediction case study further validates the dataset using a continuous test route segment, achieving a mean absolute error of 2.02~dB and a root mean square error of 2.69~dB under few-shot adaptation. WiWorld-RealData supports cross-band propagation analysis, environment-aware channel modeling, wireless digital twins, and channel foundation model research. The dataset is available at \url{https://scc.bupt.edu.cn/dataset-manage/datasets/44} and \url{https://doi.org/10.57760/sciencedb.40663}.
\end{abstract}


\section{Introduction}

The sixth-generation (6G) wireless systems are expected to evolve from connectivity-oriented communication toward sensing-assisted and intelligent wireless operation~\cite{zhang20236g,letaief2019roadmap,itur_m2160,3gpp_6g_usecases}. Therefore, networks should not only estimate instantaneous channel states, but also understand how physical environments and user states shape radio propagation over space and time~\cite{zhang2026weit}. This motivates the concept of wireless world models, which aim to establish an internal representation of the wireless environment by jointly characterizing physical scenes, user states, and propagation behavior~\cite{yu2025channelgpt,cai2026channellm}. With such a representation, wireless systems can move beyond passive channel observation toward proactive channel inference, environment-aware adaptation, and digital-twin-based network evolution~\cite{wang2024digital}.

As data-driven frameworks, wireless world models rely on the observations used for training and evaluation~\cite{zhang2019deeplearning}. Their development requires data that jointly characterize wireless propagation, environment states, and user states, while explicitly associating these observations. Such data should exhibit three key properties:
realism, multimodal completeness, and sample-level correspondence. Realism requires the preservation of hardware impairments, blockage dynamics, and propagation variability observed in field measurements. Multimodal completeness requires complementary descriptions of scene semantics, geometric structures, dynamic objects, and spatial trajectories. Sample-level correspondence requires these heterogeneous observations to be associated with channel responses under a common
spatiotemporal reference, enabling models to learn generalizable
environment--channel relationships~\cite{zhang20236g,wang2024digital,ao2025buptsounder}.

\begin{table*}[t]
    \centering
    \caption{Comparison of Representative Channel--Environment Datasets}
    \label{tab:dataset_comparison}
    \scriptsize
    \renewcommand{\arraystretch}{1.10}
    \setlength{\tabcolsep}{1.3pt}

    \begin{tabularx}{\textwidth}{
        @{}
        >{\raggedright\arraybackslash}p{0.105\textwidth}|
        >{\raggedright\arraybackslash}p{0.045\textwidth}|
        >{\hsize=1.00\hsize\linewidth=\hsize
          \raggedright\arraybackslash}X|
        >{\raggedright\arraybackslash}p{0.095\textwidth}|
        >{\hsize=1.20\hsize\linewidth=\hsize
          \raggedright\arraybackslash}X|
        >{\raggedright\arraybackslash}p{0.100\textwidth}|
        >{\raggedright\arraybackslash}p{0.095\textwidth}|
        >{\hsize=0.80\hsize\linewidth=\hsize
          \raggedright\arraybackslash}X
        @{}
    }
        \hline
        \textbf{Dataset}
        & \textbf{Source}
        & \textbf{Channel}
        & \textbf{Freq.}
        & \textbf{Environment Data}
        & \makecell[tl]{\textbf{Time}\\\textbf{Alignment}}
        & \textbf{Scale}
        & \makecell[tl]{\textbf{Sample}\\\textbf{Key}} \\
        \hline

        DeepMIMO~\cite{alkhateeb2019deepmimo}
        & Sim.
        & CFR/CSI
        & 60~GHz
        & 3D scene + user positions
        & --
        & 1.18M users
        & Location ID \\
        \hline

        WAIR-D~\cite{huangfu2022waird}
        & Sim.
        & CFR
        & \makecell[tl]{2.6/6/28/60/\\100~GHz}
        & OSM maps + BS/UE positions
        & --
        & 2.5M links
        & Link ID \\
        \hline

        CKMImageNet~\cite{wu2024ckmimagenet}
        & Sim.
        & Path loss/delay/AoA/AoD
        & 28~GHz
        & Environment maps + materials
        & --
        & $>$10k maps
        & Grid location \\
        \hline

        SynthSoM~\cite{cheng2025synthsom}
        & Sim.
        & Path loss + channel matrices
        & \makecell[tl]{Sub-6/mmWave}
        & RGB/depth + LiDAR + radar
        & Snapshot-level
        & 743.8~GB
        & Snapshot ID \\
        \hline

        Multimodal-Wireless~\cite{mao2025multimodalwireless}
        & Sim.
        & CIR/CSI
        & \makecell[tl]{4.9/28GHz}
        & Multi-view RGB/depth + LiDAR + radar + GNSS/IMU
        & \makecell[tl]{Frame-level;10~ms}
        & $\sim$160k frames
        & Frame ID \\
        \hline

        DeepSense~6G~\cite{alkhateeb2023deepsense6g}
        & Meas.
        & Beam power
        & 60~GHz
        & RGB + 2D/3D LiDAR + radar + GPS
        & UTC timestamp
        & $>$1M samples
        & Data group \\
        \hline

        \textbf{WiWorld-RealData}
        & \textbf{Meas.}
        & \textbf{CIR}
        & \makecell[tl]{\textbf{3.7/6.775~GHz}}
        & \textbf{Multi-view/panoramic images + LiDAR + radar + GNSS}
        & \makecell[tl]{\textbf{CIR-centered;}\\
          \textbf{offsets recorded}}
        & \makecell[tl]{\textbf{10~TB;}\\
          \textbf{$>$50k aligned}}
        & \makecell[tl]{\textbf{Unified}\\\textbf{sample ID}} \\
        \hline
    \end{tabularx}

    \begin{minipage}{\textwidth}
        \scriptsize
        \raggedright
        \emph{Note:} Dataset scales use different units and are not
        directly comparable.
    \end{minipage}
\end{table*}

Existing channel--environment datasets can be broadly categorized as simulation-based or measurement-based. Simulation-based datasets, including DeepMIMO, WAIR-D, CKMImageNet, SynthSoM, and Multimodal-Wireless, generate large-scale and controllable samples using ray tracing, maps, or simulated environments, and provide channel parameters, radio maps, and multimodal environment observations~\cite{alkhateeb2019deepmimo,huangfu2022waird,wu2024ckmimagenet,cheng2025synthsom,mao2025multimodalwireless}. However, their propagation processes are determined by predefined scene geometries, material properties, and propagation assumptions, which limits their ability to reproduce hardware impairments, blockage dynamics, and propagation variability observed in field measurements. Among measurement-based datasets, DeepSense~6G combines real images, LiDAR, radar, positioning information, and wireless measurements for sensing-assisted communication tasks~\cite{alkhateeb2023deepsense6g}. Its wireless data, however, mainly consist of beam-power measurements designed for beam selection and related prediction tasks, rather than channel responses for general propagation modeling. Existing datasets therefore remain limited in at least one of three aspects: real-world propagation fidelity, reusable channel representation, or sample-level associations among environment observations, user states, and channel responses. These differences are summarized in Table~\ref{tab:dataset_comparison}.

To address these limitations, this paper presents Wireless world (WiWorld)-RealData, a real-world multimodal channel--environment dataset for WWM research. Constructed from continuous outdoor measurements, the dataset jointly captures channel impulse responses (CIRs), multimodal environment observations, and transceiver positions. The sensing modalities characterize scene semantics, three-dimensional structures, dynamic objects, and spatial trajectories. A CIR-centered organization and unified metadata associate the heterogeneous records at the sample level while retaining timestamps, time offsets, and quality information. An environment-assisted path-loss prediction case study with few-shot adaptation further evaluates the utility of the aligned data for learning environment--channel relationships.

\section{Construction of WiWorld-RealData}

This section presents the construction of WiWorld-RealData from three perspectives: system configuration, route-level measurement design, and sample-level dataset organization.

\subsection{Measurement System and Configuration}

\begin{figure}
    \centering
    \includegraphics[width=0.95\linewidth]{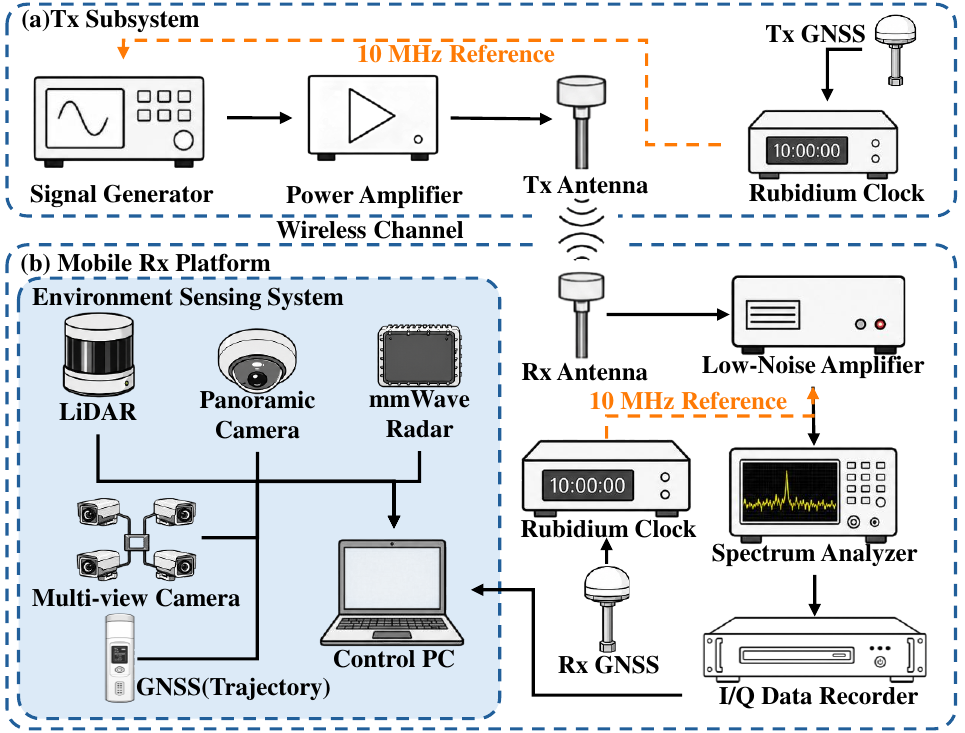}
    \caption{Architecture of the joint channel–environment measurement system.}
    \label{fig:measure}
\end{figure}

WiWorld-RealData was collected using an upgraded BUPTSounder Pro joint channel--environment acquisition system~\cite{ao2025buptsounder}, as shown in Fig.~\ref{fig:measure}. The system comprises a wideband channel measurement subsystem, a multimodal environment sensing subsystem, and a time synchronization and control subsystem. It supports continuous field measurements with a fixed transmitter (Tx) and a mobile receiver (Rx).

For channel sounding, the Tx radiates a wideband PN11 sounding signal, while the mobile Rx continuously samples and records complex in-phase/quadrature (I/Q) data. Either a horn or an omnidirectional antenna is used at the Tx, while an omnidirectional antenna is used at the Rx. The recorded I/Q data are correlated with a calibrated reference sequence to obtain wideband channel impulse responses (CIRs). Global navigation satellite system (GNSS) receivers and rubidium clocks at the Tx and Rx provide a common timing reference. The main measurement configurations, antenna models, and measured gains are summarized in Table~\ref{tab:measurement_config}.

The environment sensing subsystem mounted on the mobile Rx platform integrates four network cameras, a $360^{\circ}$ panoramic camera, a light detection and ranging (LiDAR) sensor, a millimeter-wave (mmWave) radar, and a GNSS receiver. All sensing devices are triggered by a unified control script, and their timestamps are retained. Panoramic images, LiDAR point clouds, and GNSS records are also collected at the fixed Tx location to provide Tx-side environment observations.

\vspace{2mm}
\begin{table}[t]
    \centering
    \caption{Main Measurement Configurations of WiWorld-RealData}
    \label{tab:measurement_config}
    \footnotesize
    \renewcommand{\arraystretch}{1.0}
    \setlength{\tabcolsep}{3pt}
    \renewcommand{\tabularxcolumn}[1]{m{#1}}

    \begin{tabularx}{\columnwidth}{
        >{\raggedright\arraybackslash}m{0.40\columnwidth}|
        >{\centering\arraybackslash}X|
        >{\centering\arraybackslash}X
    }
        \hline
        \textbf{Parameter}
        & \textbf{3.7~GHz}
        & \textbf{6.775~GHz} \\
        \hline

        Bandwidth
        & 200~MHz
        & 600~MHz \\
        \hline

        Sampling rate
        & 200~MSamples/s
        & 600~MSamples/s \\
        \hline

        Delay resolution
        & 5.00~ns
        & 1.67~ns \\
        \hline

        Horn antenna gain
        & 10.21~dBi
        & 10.20~dBi \\
        \hline

        Omnidirectional gain
        & 0.95~dBi
        & 0.99~dBi \\
        \hline

        Tx/Rx heights
        & \multicolumn{2}{c}{20.5~m / 2.5~m} \\
        \hline

        Tx--Rx distance
        & \multicolumn{2}{c}{5--375~m} \\
        \hline

        Rx speed
        & \multicolumn{2}{c}{1--1.4~m/s} \\
        \hline

        Sensing rates
        & \multicolumn{2}{c}{
            \makecell[c]{
                Cameras/panorama: 10~fps; LiDAR: 10~Hz;\\
                radar/GNSS: 20~Hz
            }
        } \\
        \hline
    \end{tabularx}
\end{table}

\subsection{Measurement Scenario and Routes}

The measurements are conducted in the outdoor areas of the Haidian Campus of Beijing University of Posts and Telecommunications. The measurement area covers building-lined roads, open spaces, vegetated road sections, parking areas, and shared pedestrian--vehicle zones, representing a typical campus-urban environment. These environment elements introduce blockage, reflection, scattering, and time-varying disturbances, providing diverse propagation conditions within a compact area.

The Tx is fixed at an elevated position on a campus building, while the Rx moves continuously along predefined routes on a mobile platform. The route design covers continuous transitions among line-of-sight (LoS), obstructed line-of-sight (OLoS), and non-line-of-sight (NLoS) conditions. These transitions produce spatial variations in path loss and multipath characteristics, enabling joint observation of environment changes and channel evolution.

\begin{figure}[t]
    \centering
    \includegraphics[width=0.45\textwidth]{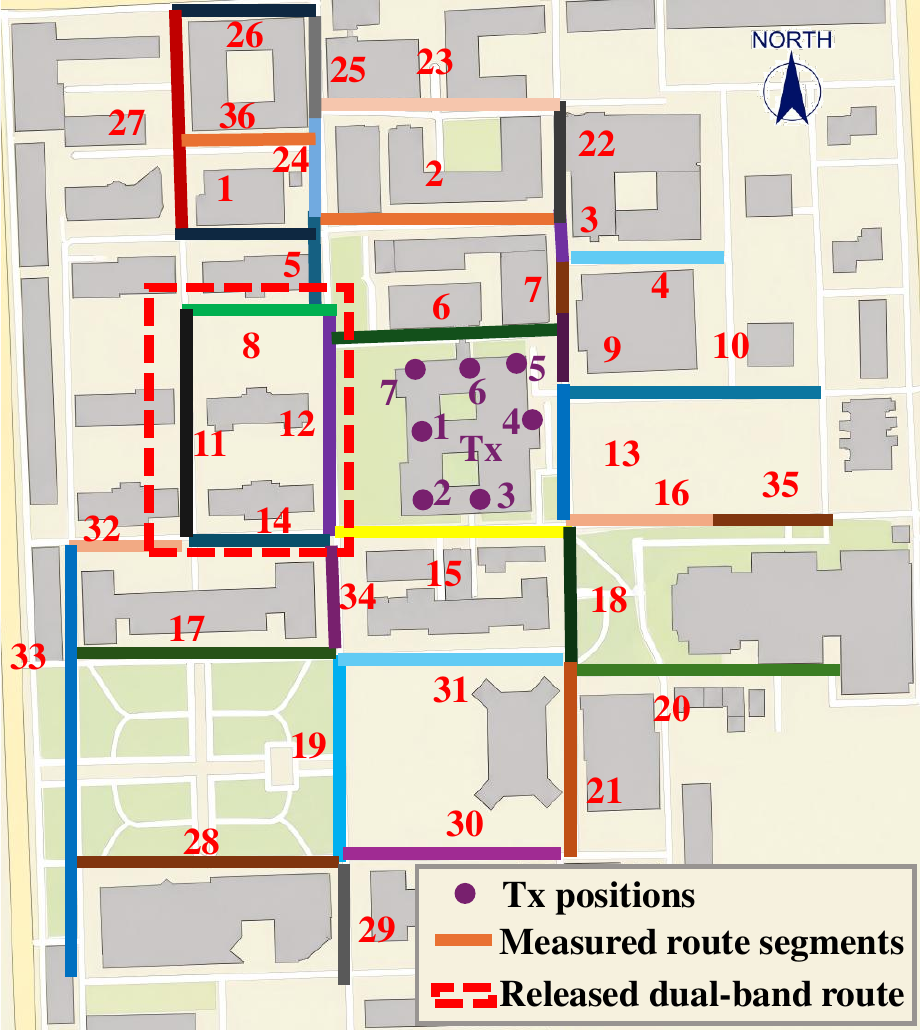}
    \caption{Measurement route segments and Tx locations on the BUPT Haidian Campus. The highlighted region indicates the released dual-band route. Different colors are used only to distinguish adjacent route segments.}
    \label{fig:measurement_routes}
\end{figure}

The measurement campaign comprises 15 continuous routes consisting of 36 outdoor route segments, as shown in Fig.~\ref{fig:measurement_routes}. These segments cover different relative Tx--Rx geometries, antenna orientations, road layouts, and blockage conditions. Channel and environment observations are continuously acquired along each route. The current public release includes one route at 3.7~GHz and one route at 6.775~GHz within the highlighted region, together with their corresponding multimodal environment records.

\subsection{Dataset Organization and Sample Alignment}

\begin{figure*}[t]
    \centering
    \includegraphics[width=\textwidth]{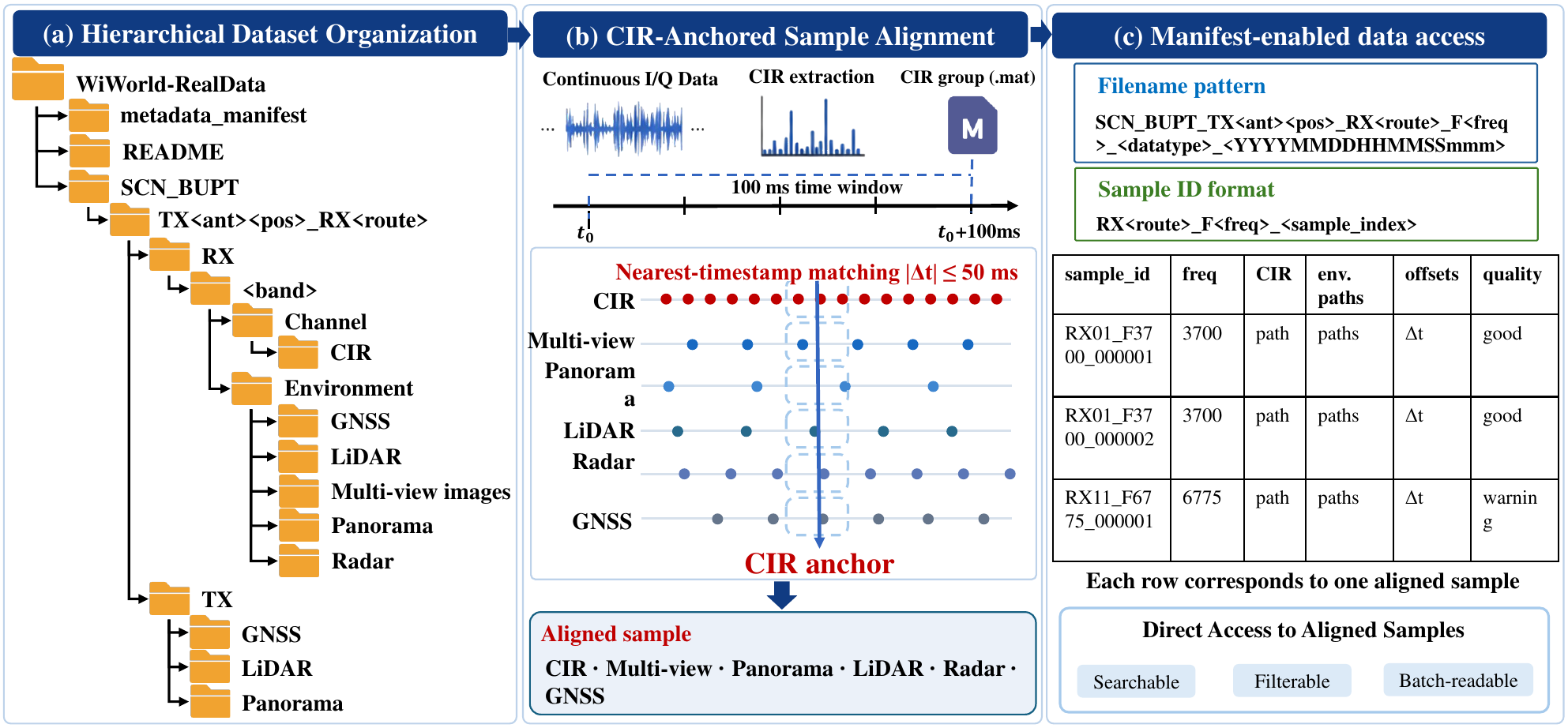}
    \caption{Dataset organization, CIR-anchored sample alignment, and manifest-enabled data access in WiWorld-RealData.}
    \label{fig:dataset_organization}
\end{figure*}

WiWorld-RealData follows a scenario--route--frequency--sample hierarchy, as illustrated in Fig.~\ref{fig:dataset_organization}. CIRs and multimodal environment sensing data are organized in modality-specific directories and indexed through a unified metadata manifest. The complete measurement campaign generated approximately 10~TB of data, including continuously recorded wideband I/Q measurements, more than 200{,}000 multi-view images, approximately 50{,}000 panoramic images, 50{,}000 LiDAR point clouds, 50{,}000 mmWave radar records, and 100{,}000 GNSS records.

The continuous wideband I/Q measurements are processed into complex CIRs, and the timestamp of each CIR is derived from the measurement start time, sampling rate, and global sample index. CIRs whose start times fall within the same 100-ms window are grouped into one \texttt{.mat} file, while their precise timestamps and global sample indices are retained. Each CIR group is indexed by its 100-ms window, and the original timestamps of the multimodal environment sensing records are mapped to the same 100-ms time grid without discarding their precise acquisition times. Each CIR file then serves as the anchor and is matched with the temporally nearest record from each sensing modality when the absolute time difference does not exceed 50~ms. The actual time differences are recorded, while unmatched file-path and time-offset fields are marked as \texttt{NAN}.

Each row of the unified metadata manifest represents one aligned channel--environment sample and records its sample identifier, Tx and Rx identifiers, center frequency, reference timestamp, file paths, time offsets, and quality flags. Data files follow the naming convention \texttt{SCN\_BUPT\_TX<ant><pos>\_RX<route>\_F<freq>\_}\
\texttt{<datatype>\_<YYYYMMDDHHMMSSmmm>}. Each aligned sample is assigned a unique identifier following the pattern \texttt{RX<route>\_F<freq>\_<sample\_index>}, with the sample index increasing independently within each Rx--frequency combination. Faces and license plates in the released images are anonymized.

\section{Advantages of WiWorld-RealData}

WiWorld-RealData is distinguished by real-world field acquisition, rich multimodal environment sensing, dual-band channel measurements, continuous propagation-state transitions, sample-level temporal alignment, large-scale data volume, and standardized data organization. Compared with the representative datasets summarized in Table~\ref{tab:dataset_comparison}, WiWorld-RealData provides four principal advantages for environment--channel learning: real-world measurements, multimodal environment representation, dual-band CIRs, and large-scale sample-level alignment.

\textbf{Real-World Measurements.} Simulation-based datasets derive channel responses from predefined geometries, material properties, and propagation mechanisms, and therefore inherit the limitations imposed by the underlying simulator. WiWorld-RealData instead records continuous channel evolution through field measurements, capturing propagation effects induced by actual blockage, reflection, diffraction, and scattering. Continuous transitions among line-of-sight (LoS), obstructed line-of-sight (OLoS), and non-line-of-sight (NLoS) conditions further provide measured propagation sequences rather than isolated channel states. These properties support empirical channel characterization, blockage and propagation-state analysis, and fidelity assessment of digital-twin channel models.

\begin{figure*}[t]
    \centering
    \includegraphics[width=0.98\textwidth]{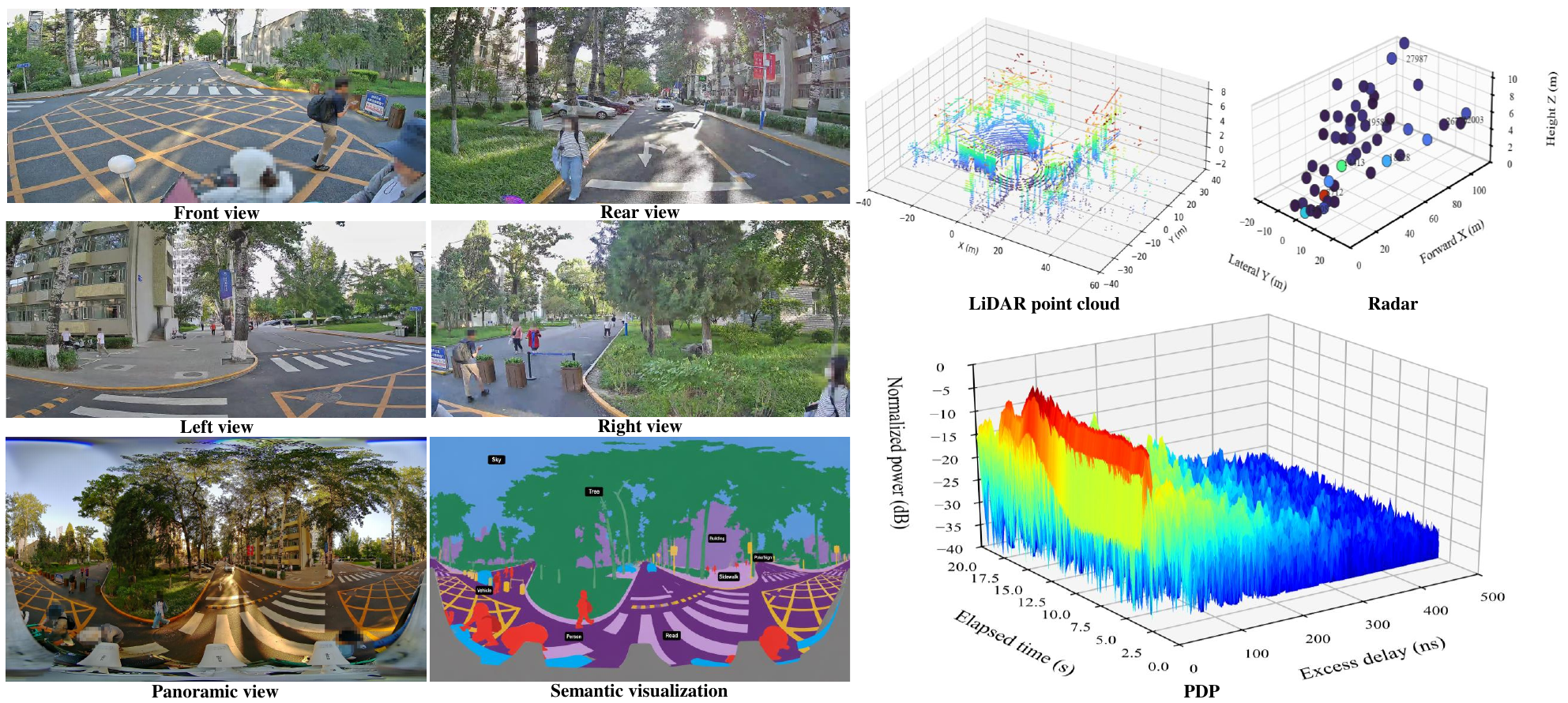}
    \caption{Representative aligned multimodal sample in WiWorld-RealData. Semantic results are shown for visualization only.}
    \label{fig:multimodal_data}
\end{figure*}

\textbf{Multimodal Environment Representation.} A single sensing modality captures only one aspect of the propagation environment. Visual observations provide semantic context, LiDAR resolves three-dimensional geometry, mmWave radar captures dynamic-object information, and GNSS records positions and trajectories. WiWorld-RealData associates these complementary observations with measured CIRs within aligned samples, linking semantic, geometric, dynamic, and positional information to the corresponding channel responses. This association supports environment-assisted channel prediction, blockage and propagation-state identification, multipath interpretation, and joint environment--channel modeling. Representative aligned multimodal observations and the corresponding PDP evolution are shown in Fig.~\ref{fig:multimodal_data}. The PDP is visualized over a 20-s interval to illustrate the temporal evolution of the measured channel, while the displayed multimodal observations correspond to the channel state at $t=10$~s.

\begin{figure}[t]
    \centering
    \includegraphics[width=1\linewidth]{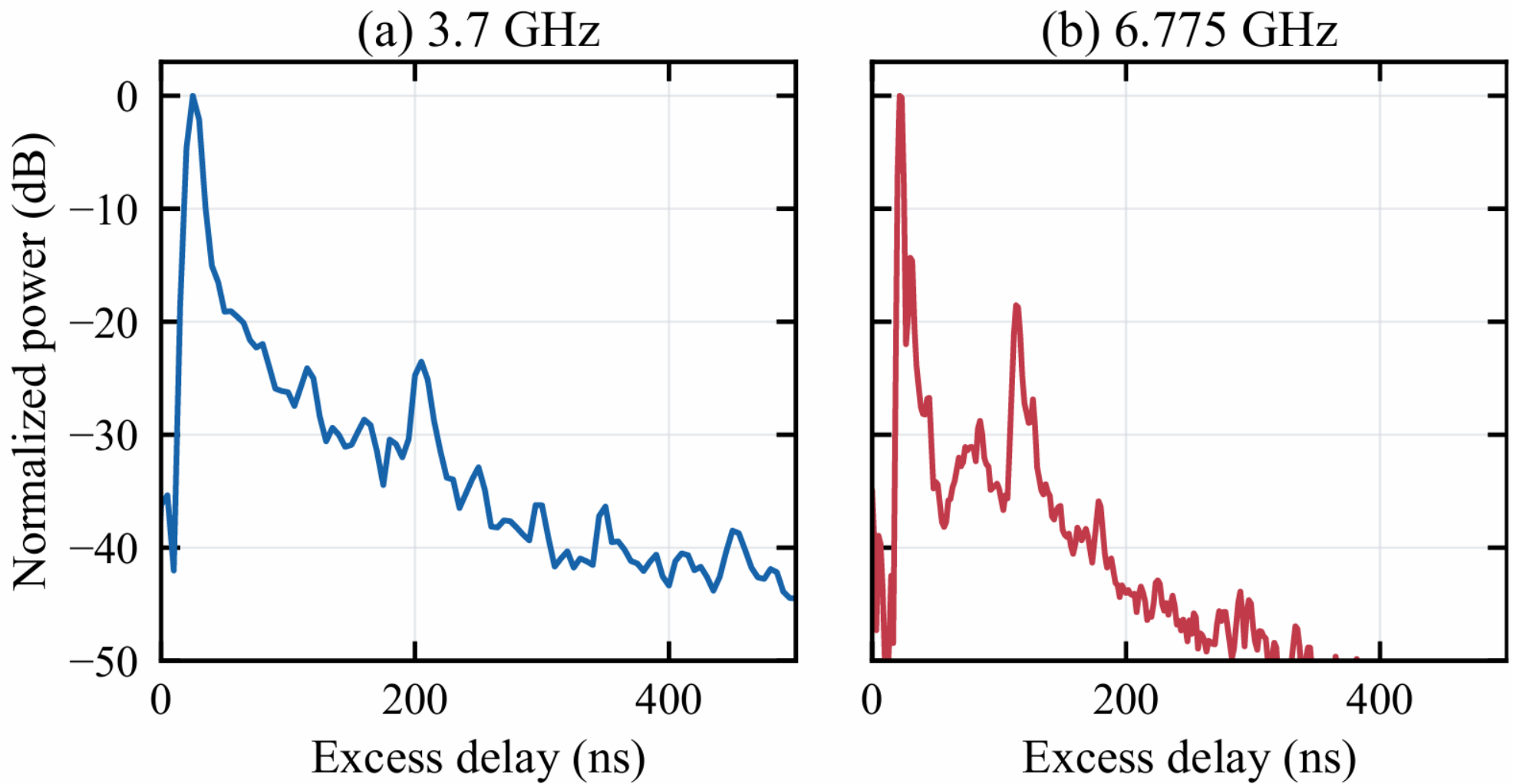}
    \caption{Representative normalized PDPs at (a) 3.7~GHz and (b) 6.775~GHz.}
    \label{fig:dual_band_comparison}
\end{figure}

\textbf{Dual-Band CIRs.} Among the representative datasets in Table~\ref{tab:dataset_comparison}, multi-frequency channel data are predominantly simulation-based, whereas measured multimodal datasets rarely provide complete wideband CIRs at multiple frequencies. WiWorld-RealData provides measured complex CIRs at 3.7 and 6.775~GHz under the same route geometry, covering a widely used 5G mid-band frequency and an upper-6-GHz frequency of interest for future wireless systems~\cite{xu2026nearfield}. As shown in Fig.~\ref{fig:dual_band_comparison}, the PDPs at both frequencies preserve resolvable multipath components and exhibit different excess-delay distributions, demonstrating the dataset's capability to capture frequency-dependent propagation characteristics. These route-consistent, delay-resolved measurements support cross-frequency propagation analysis, frequency extrapolation, and multi-band channel prediction.

\textbf{Large-Scale Sample-Level Alignment.} Channel and environment measurements acquired by independent devices form heterogeneous data streams with different sampling rates and recording granularities and therefore do not directly constitute one-to-one learning samples. WiWorld-RealData organizes these measurements into more than 50{,}000 traceable channel--environment samples through CIR-anchored temporal matching and a unified metadata manifest. The explicit cross-modal correspondence enables direct sample retrieval, quality-based filtering, and batch loading, reducing repeated cross-modal matching and supporting reproducible large-scale environment--channel learning and model evaluation.

\section{Case Study: Environment-Assisted Path-Loss Prediction}
\label{sec:case_study}

This section evaluates the utility of WiWorld-RealData through an environment-assisted path-loss prediction task. A regression model is trained using the aligned environment observations and CIR-derived path-loss labels, and its performance is evaluated on a spatially continuous target route segment after adaptation with a limited number of target samples.

\subsection{Task Definition and Data Label}

The task is formulated as a regression problem. For the $i$th aligned sample, the model input is defined as
\begin{equation}
    \mathbf{x}_i =
    \left\{
    \mathbf{I}_i^{\mathrm{mv}},
    \mathbf{I}_i^{\mathrm{pan}},
    \mathbf{P}_i^{\mathrm{LiDAR}},
    \mathbf{p}_i^{\mathrm{Tx}},
    \mathbf{p}_i^{\mathrm{Rx}}
    \right\},
    \label{eq:model_input}
\end{equation}
where $\mathbf{I}_i^{\mathrm{mv}}$ denotes the four-view images, $\mathbf{I}_i^{\mathrm{pan}}$ denotes the panoramic image, $\mathbf{P}_i^{\mathrm{LiDAR}}$ denotes the LiDAR point cloud, and $\mathbf{p}_i^{\mathrm{Tx}}$ and $\mathbf{p}_i^{\mathrm{Rx}}$ denote the Tx and Rx positions, respectively. The path-loss estimate for the $i$th sample is given by
\begin{equation}
    \widehat{\mathrm{PL}}_i = f_{\boldsymbol{\theta}}\left(\mathbf{x}_i\right),
    \label{eq:path_loss_prediction}
\end{equation}
where $f_{\boldsymbol{\theta}}(\cdot)$ denotes the regression model parameterized by $\boldsymbol{\theta}$.

The supervision label is derived from the calibrated CIR data measured at 3.7~GHz. Each aligned sample is anchored by one 100-ms CIR file containing $N_i$ short-term CIR snapshots. Let $h_i[n,\ell]$ denote the calibrated complex channel response of the $n$th snapshot at the $\ell$th delay bin. The path-loss label is calculated as
\begin{equation}
    \mathrm{PL}_i =
    -10\log_{10}
    \left(
    \frac{1}{N_i}
    \sum_{n=1}^{N_i}
    \sum_{\ell\in\Omega_i}
    \left|h_i[n,\ell]\right|^2
    \right),
    \label{eq:path_loss_label}
\end{equation}
where $\Omega_i$ denotes the set of valid delay bins used for power integration.

\subsection{Validation Model and Experimental Configuration}

A multimodal regression model composed of established feature encoders is adopted to evaluate the usability of the aligned channel--environment samples. ResNet-34~\cite{he2016deep} extracts visual features from the four-view and panoramic images, PointNet++~\cite{qi2017pointnetplusplus} encodes the LiDAR point cloud, and a multilayer perceptron encodes the Tx and Rx positions. The resulting features are projected into modality tokens and processed by a GPT-2-based fusion module~\cite{radford2019language} to model cross-modal dependencies. A regression head subsequently maps the fused representation to the predicted path loss. The model is optimized using the mean squared error loss.

The experiment uses 2,889 aligned samples collected at 3.7~GHz under the same Tx--Rx configuration. Samples 150--750 in the acquisition sequence form a spatially continuous target-domain segment, while all remaining samples constitute the source domain. The source-domain samples are divided into training and validation sets at a ratio of 9:1. Within the target-domain segment, 20 labeled samples, corresponding to approximately 3.3\% of the target samples, are used for model adaptation, and the remaining target samples are used for evaluation.

Training is conducted in two stages. The model is first trained on the source-domain training set for at most 500 epochs, and the parameters achieving the lowest source-domain validation loss are retained. Early stopping is applied when the validation loss does not improve for 30 consecutive epochs. The resulting model is then adapted using the 20 labeled target-domain samples before being evaluated on the remaining continuous target segment.

\subsection{Results and Dataset Utility}

\begin{figure}[t]
    \centering
    \includegraphics[width=\linewidth]{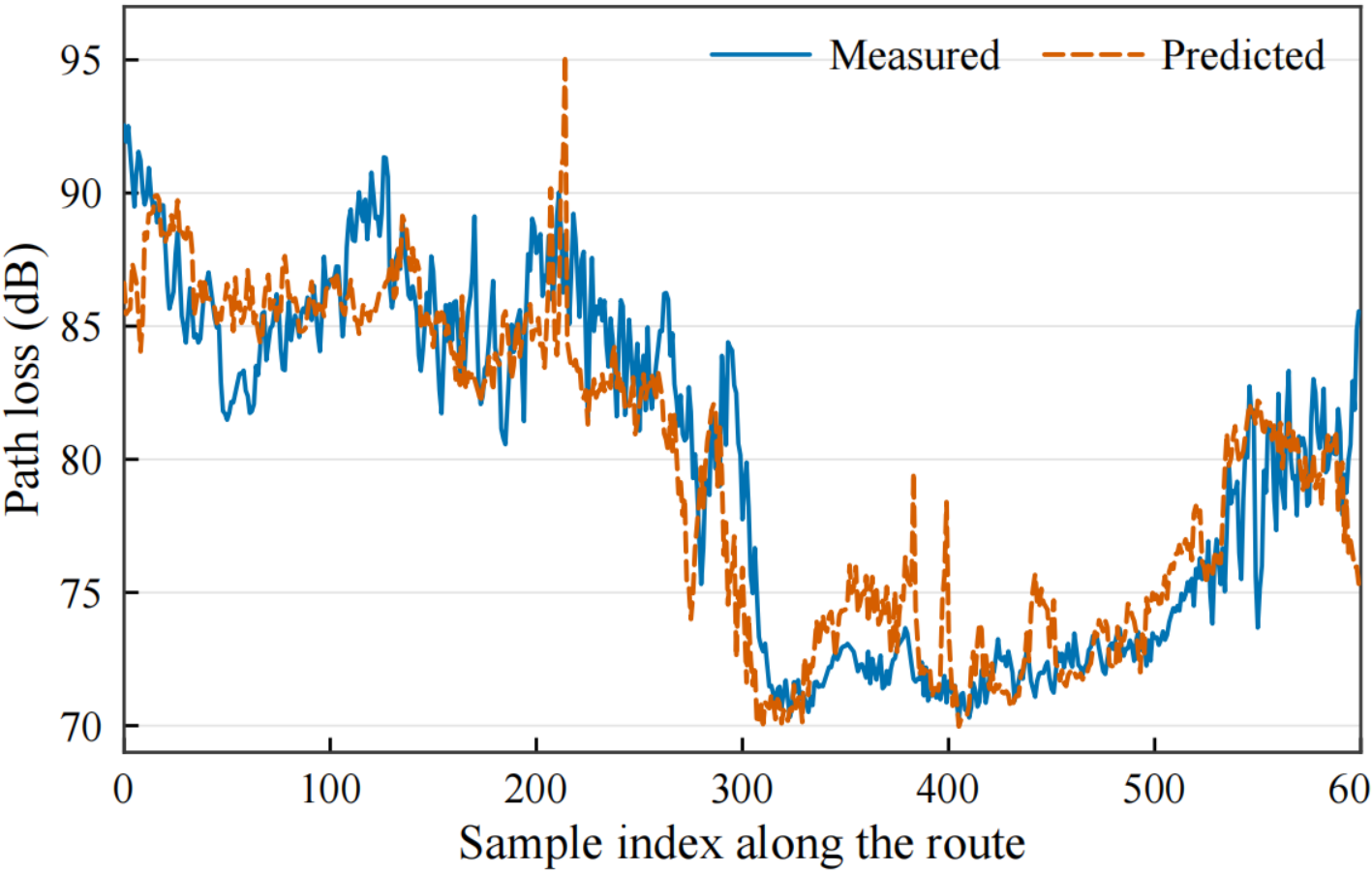}
    \caption{Measured and predicted path loss along the continuous target route segment after adaptation with 20 labeled target samples.}
    \label{fig:path_loss_results}
\end{figure}

After adaptation with 20 labeled target-domain samples, the model achieves a mean absolute error (MAE) of 2.02~dB, a root mean square error (RMSE) of 2.69~dB, and a median absolute error of 1.55~dB. As shown in Fig.~\ref{fig:path_loss_results}, the predicted values follow the principal variations of the measured path loss along the continuous target segment. This agreement indicates that the aligned visual, geometric, and positional observations contain information relevant to large-scale channel variation. Larger errors occur primarily near road corners, blockage boundaries, and rapidly changing propagation regions, where the relative contributions of multipath components vary over short spatial intervals.

The ablation results in Table~\ref{tab:ablation} show that the complete multimodal input achieves the lowest MAE and RMSE. Removing the multi-view images results in a slight performance degradation, whereas removing the panoramic image produces a larger error increase, indicating that the two visual representations provide complementary environment information.

\begin{table}[t]
    \caption{Ablation Results with Different Environment Modalities}
    \label{tab:ablation}
    \centering
    \footnotesize
    \setlength{\tabcolsep}{4pt}
    \renewcommand{\arraystretch}{1.10}

    \begin{tabular}{l|c|c}
        \hline
        \textbf{Input Modalities}
        & \textbf{MAE (dB)}
        & \textbf{RMSE (dB)} \\
        \hline

        Multi-view + Panorama + LiDAR + GNSS
        & \textbf{2.02}
        & \textbf{2.69} \\
        \hline

        Panorama + LiDAR + GNSS
        & 2.06
        & 2.78 \\
        \hline

        Multi-view + LiDAR + GNSS
        & 2.45
        & 3.50 \\
        \hline
    \end{tabular}
\end{table}

\section{Conclusion}

This paper presents WiWorld-RealData, a real-world multi-band channel and multi-modal environment sensing dataset for 6G wireless world model research. The dataset provides measured CIRs at 3.7 and 6.775~GHz together with aligned environment observations, while unified timestamps, sample identifiers, and metadata establish explicit sample-level correspondences across heterogeneous modalities. An environment-assisted path-loss prediction case study achieves an MAE of 2.02~dB and an RMSE of 2.69~dB under few-shot adaptation, validating the utility of the dataset for environment--channel relationship learning. WiWorld-RealData provides a reusable data foundation for cross-band propagation analysis, environment-aware channel modeling, wireless digital twins, and channel foundation model research.
\bibliographystyle{IEEEtran}
\bibliography{WiWorld-Real}

@article{zhang20236g,
  title={{6G} Channel Modeling: Requirement, Measurement, Methodology and Simulator},
  author={Zhang, Jianhua and Lin, Jiaxin and Tang, Pan and others},
  journal={arXiv preprint arXiv:2305.16616},
  year={2023},
  doi={10.48550/arXiv.2305.16616}
}

@article{wang2024digital,
  title={Digital Twin Channel for {6G}: Concepts, Architectures
         and Potential Applications},
  author={Wang, Heng and Zhang, Jianhua and Nie, Gaofeng and others},
  journal={IEEE Communications Magazine},
  volume={63},
  number={3},
  pages={24-30},
  year={2025},
  doi={10.1109/MCOM.001.2400213}
}

@inproceedings{ao2025buptsounder,
  title={{BUPTSounder Pro}: A Multi-Modal Environment-Channel Joint
         Data Acquisition System for {6G} Digital Twin Channel},
  author={Ao, B. and Yu, L. and Zhang, Y. and others},
  booktitle={2025 {IEEE} {GLOBECOM} Workshops ({GC Wkshps})},
  pages={1196-1201},
  year={2025}
}

@article{alkhateeb2019deepmimo,
  title={{DeepMIMO}: A Generic Deep Learning Dataset for Millimeter
         Wave and Massive {MIMO} Applications},
  author={Alkhateeb, Ahmed},
  journal={arXiv preprint arXiv:1902.06435},
  year={2019},
  doi={10.48550/arXiv.1902.06435}
}

@article{huangfu2022waird,
  title={{WAIR-D}: Wireless {AI} Research Dataset},
  author={Huangfu, Yourui and Wang, Jian and Dai, Shengchen and others},
  journal={arXiv preprint arXiv:2212.02159},
  year={2022},
  doi={10.48550/arXiv.2212.02159}
}

@inproceedings{wu2024ckmimagenet,
  title={{CKMImageNet}: A Comprehensive Dataset to Enable Channel
         Knowledge Map Construction via Computer Vision},
  author={Wu, Di and Wu, Zijian and Qiu, Yuelong and others},
  booktitle={2024 {IEEE/CIC} International Conference on Communications
             in China ({ICCC} Workshops)},
  pages={114-119},
  year={2024},
  doi={10.1109/ICCCWorkshops62562.2024.10693754}
}

@article{cheng2025synthsom,
  title={{SynthSoM}: A Synthetic Intelligent Multi-Modal
         Sensing-Communication Dataset for Synesthesia of Machines ({SoM})},
  author={Cheng, Xiang and Huang, Ziwei and Yu, Yong and others},
  journal={Scientific Data},
  volume={12},
  pages={819},
  year={2025},
  doi={10.1038/s41597-025-05065-x}
}

@article{mao2025multimodalwireless,
  title={{Multimodal-Wireless}: A Large-Scale Dataset for Sensing
         and Communication},
  author={Mao, Tianhao and Liang, Le and Yang, Jie and others},
  journal={arXiv preprint arXiv:2511.03220},
  year={2025},
  doi={10.48550/arXiv.2511.03220}
}

@article{alkhateeb2023deepsense6g,
  title={{DeepSense {6G}}: A Large-Scale Real-World Multi-Modal
         Sensing and Communication Dataset},
  author={Alkhateeb, Ahmed and Charan, Gouranga and Osman, Tawfik and others},
  journal={IEEE Communications Magazine},
  volume={61},
  number={9},
  pages={122-128},
  year={2023},
  doi={10.1109/MCOM.006.2200730}
}

@inproceedings{he2016deep,
  title={Deep Residual Learning for Image Recognition},
  author={He, Kaiming and Zhang, Xiangyu and Ren, Shaoqing and others},
  booktitle={Proceedings of the {IEEE} Conference on Computer Vision
             and Pattern Recognition ({CVPR})},
  pages={770-778},
  year={2016},
  doi={10.1109/CVPR.2016.90}
}

@inproceedings{qi2017pointnetplusplus,
  title={{PointNet++}: Deep Hierarchical Feature Learning on Point
         Sets in a Metric Space},
  author={Qi, Charles Ruizhongtai and Yi, Li and Su, Hao and others},
  booktitle={Advances in Neural Information Processing Systems},
  volume={30},
  pages={5099-5108},
  year={2017}
}

@article{yu2025channelgpt,
  title={{ChannelGPT}: A Large Model Toward Real-World Channel
         Foundation Model for {6G} Environment Intelligence Communication},
  author={Yu, Li and Shi, Lianzheng and Zhang, Jianhua and others},
  journal={IEEE Communications Magazine},
  volume={63},
  number={10},
  pages={68-74},
  year={2025},
  doi={10.1109/MCOM.001.2400780}
}

@techreport{radford2019language,
  title={Language Models Are Unsupervised Multitask Learners},
  author={Radford, Alec and Wu, Jeffrey and Child, Rewon and others},
  institution={OpenAI},
  year={2019},
  url={https://cdn.openai.com/better-language-models/language_models_are_unsupervised_multitask_learners.pdf}
}

@techreport{itur_m2160,
  author      = {{ITU-R}},
  title       = {Framework and Overall Objectives of the Future Development of IMT for 2030 and Beyond},
  institution = {International Telecommunication Union Radiocommunication Sector},
  number      = {Recommendation ITU-R M.2160-0},
  year        = {2023},
  month       = nov
}

@techreport{3gpp_6g_usecases,
  author      = {{3GPP}},
  title       = {Study on 6G Use Cases and Service Requirements},
  institution = {3rd Generation Partnership Project},
  number      = {3GPP TR 22.870, Release 20},
  year        = {2026}
}

@article{zhang2026weit,
  title={Wireless Environmental Information Theory: A New Paradigm Toward {6G} Online and Proactive Environment Intelligence Communication},
  author={Zhang, Jianhua and Yu, Li and Liu, Shaoyi and others},
  journal={Engineering},
  volume={56},
  pages={186-200},
  year={2026},
  doi={10.1016/j.eng.2025.07.028}
}

@article{cai2026channellm,
  author  = {Cai, Yichen and Qiu, Yuelong and Zhang, Jianhua and others},
  title   = {Paradigm Shift from Statistical Channel Modeling to Digital Twin Prediction: An Environment-Generalizable {ChannelLM} for {6G} {AI}-Enabled Air Interface},
  journal = {arXiv preprint arXiv:2604.18021},
  year    = {2026}
}

@article{xu2026nearfield,
  author  = {Xu, Huixin and Zhang, Jianhua and Tang, Pan and others},
  title   = {Near-Field Propagation and Spatial Non-Stationarity Channel Model for {6--24 GHz} ({FR3}) Extremely Large-Scale {MIMO}: Adopted by {3GPP} for {6G}},
  journal = {IEEE Journal on Selected Areas in Communications},
  volume  = {44},
  pages   = {3201-3218},
  year    = {2026},
  doi     = {10.1109/JSAC.2026.3650888}
}

@article{letaief2019roadmap,
  author  = {Letaief, Khaled B. and Chen, Wei and Shi, Yuanming and others},
  title   = {The Roadmap to {6G}: {AI} Empowered Wireless Networks},
  journal = {IEEE Communications Magazine},
  volume  = {57},
  number  = {8},
  pages   = {84--90},
  year    = {2019},
  doi     = {10.1109/MCOM.2019.1900271}
}

@article{zhang2019deeplearning,
  author  = {Zhang, Chaoyun and Patras, Paul and Haddadi, Hamed},
  title   = {Deep Learning in Mobile and Wireless Networking: A Survey},
  journal = {IEEE Communications Surveys \& Tutorials},
  volume  = {21},
  number  = {3},
  pages   = {2224-2287},
  year    = {2019},
  doi     = {10.1109/COMST.2019.2904897}
}

\end{document}